\documentclass[5p]{elsarticle}
\usepackage{lineno,dcolumn}
\linenumbers

\usepackage{amssymb}

\journal{Nuclear Physics A}

\begin{document}

\begin{frontmatter}



\title{Single Event Effect Characterization of the Mixed-Signal ASIC
Developed for CCD Camera in Space Use}


 \author[label1]{Hiroshi Nakajima\corref{cor1}}
 \author[label1]{Mari Fujikawa}
 \author[label1]{Hideki Mori}
 \author[label1]{Hiroaki Kan}
 \author[label1]{Shutaro Ueda}
 \author[label1]{Hiroko Kosugi}
 \author[label1]{Naohisa Anabuki}
 \author[label1]{Kiyoshi Hayashida}
 \author[label1]{Hiroshi Tsunemi}
 \author[label2]{John P. Doty}
 \author[label3]{Hirokazu Ikeda}
 \author[label4]{Hisashi Kitamura}
 \author[label4]{and Yukio Uchihori}
 \address[label1]{Department of Earth and Space Science, Graduate School of Science, Osaka
University, 1-1 Machikaneyama, Toyonaka, Osaka, 560-0043, Japan}
 \address[label2]{Noqsi Aerospace Ltd.,　
2822 South Nova Road, Pine, Colorado 80470, USA}
 \address[label3]{the Institute of Space and Astronautical Science,
Japan Aerospace Exploration Agency,
3-1-1 Yoshinodai, Chuo-ku, Sagamihara, Kanagawa, 252-5210, Japan}
 \address[label4]{National Institute for Radiological Sciences (NIRS),
Anagawa 4-9-1, Inage-ku, Chiba-shi, Chiba, 263-8555, Japan}

\cortext[cor1]{Further author information: (Send correspondence to H.N.)\\H.N.:
E-mail: nakajima@ess.sci.osaka-u.ac.jp, Telephone: +81 6 6850 5478}

\begin{abstract}
We present the single event effect (SEE) tolerance of a mixed-signal application-specific integrated circuit 
(ASIC) developed for a charge-coupled device camera onboard a future X-ray astronomical mission.
We adopted proton and heavy ion beams at HIMAC/NIRS in Japan. The particles with high linear energy transfer
(LET) of 57.9~$\!$MeV$\cdot$cm$^2$/mg is used to measure the single event latch-up (SEL) tolerance, which
results in a sufficiently low cross-section of $\sigma_{\rm SEL}$ $< 4.2\times10^{-11}$~$\!$cm$^2$/(Ion$\times$ASIC).
The single event upset (SEU) tolerance is estimated with various kinds of species with wide range of energy.
Taking into account that a part of the protons creates recoiled heavy ions that has higher LET than
that of the incident protons, we derived the probability of SEU event as a function of LET. Then the SEE event rate in a
low-earth orbit is estimated considering a simulation result of LET spectrum. SEL rate is below once per 49 years,
which satisfies the required latch-up tolerance. The upper limit of the SEU rate is
derived to be 1.3$\times10^{-3}$~$\!$events/sec. Although the SEU events cannot be distinguished 
from the signals of X-ray photons from astronomical objects, the derived SEU rate is below 1.3~$\!$\% of expected
non X-ray background rate of the detector and hence these events should not be a major component of the
instrumental background.
\end{abstract}


\end{frontmatter}


\section{Introduction}
\label{sec:intro}
The X-ray charge-coupled device (CCD) cameras have been functioning as the primary focal plane detector
of modern X-ray astronomy thanks to their well-balanced performances for imaging-spectroscopy
\cite{Weisskopf02,Jansen00,Mitsuda07}. 
Although we owe the imaging performance of a camera to the X-ray telescope, the energy
and timing resolutions depend heavily both on the readout noise and on the processing speed
of the front-end electronics, respectively.
Since the forthcoming camera systems will require a larger number of pixels,
a higher frame rate and a lower readout noise than those of the conventional systems,
implementing the electronics with discrete integrated circuits (ICs)
will dissipate too much power in orbit.

Recently, application-specific integrated circuits (ASICs)
have been applied to the readout systems of the detectors
including X-ray CCDs \cite{Rando04,Tajima04,Herrmann07}.
Drastically curtailed power consumption and compact size
allow us to equip many chips next to the sensors.
Hence we have developed the mixed-signal Complementary
Metal Oxide Semiconductor (CMOS) ASIC as the signal processing electronics for
CCD cameras in space use \cite{Matsuura07, Nakajima09A}.
It is characterized with not only a low-noise pre-amplifier but also
a $\Delta\Sigma$ type analog-to-digital converter ($\Delta\Sigma$ modulator), which
has noise shaping capability \cite{Inose62}.
The spectroscopic performance has been verified \cite{Nakajima09B}, which
approved the ASIC as the front-end electronics of the X-ray
CCD camera \cite{Hayashida11} onboard the ASTRO-H mission \cite{Takahashi10}.

The radiation damages to space-use ICs are classified into two types: a total ionization dose (TID)
effect and a single event effect (SEE). The former is the accumulating damage mainly due to
the geomagnetically trapped protons and electrons. The latter is the stochastic
damage that occurs when the energy deposited by a single heavy ion in the Galactic cosmic-rays
exceed a threshold defined by the linear energy transfer (LET).

SEE can further be categorized into a single event latch-up (SEL) and a single event upset (SEU).
SEL is a peculiar issue in CMOS Large Scale Integration (LSI).
Since the NMOS and PMOS transistors are formed on a single silicon wafer,
there are PNPN junctions forming parasitic thyristors \cite{Gregory73}. If the input voltage exceeds
that of the power supply momentarily due to the impingement of high energy particle,
the thyristor is switched on and the excessive current continues to flow unless
the power is shutdown. Finally, the LSI may be broken due to the overheat.
On the other hand, SEU is generally represented by the bit inversion in the circuits
such as memories and flip-flops. Despite the fact that SEU events are not critically
harmful to the LSI, it can fudges up false signals.

\citet{Nakajima11} performed the TID test by irradiating our ASICs with a proton beam
and verified that the gain and the input equivalent noise
were not influenced up to 140~krad. The simulated dose rate in the low earth orbit (LEO)
of ASTRO-H is 1~krad/yr, which promises no significant accumulating damage during the required
mission lifetime of three years. This paper reports the SEE tolerance of the ASIC measured
with protons and heavy ions.

After the description of our ASIC and the specification of the SEE test
in Section~\ref{sec:test}, we report the results (Section~\ref{sec:results})
and the expected SEE rate in the LEO (Section~\ref{sec:discussion}), followed
by summary (Section~\ref{sec:summary}).
Indicated errors below mean 90~\% confidence level, unless otherwise mentioned.

\section{Specification of the SEE Test}
\label{sec:test}

\subsection{Specification of the ASIC}
\label{ssec:asic}

The detailed specification of our ASIC  (hereafter we call it MND02) is described in
\citet{Nakajima11} and references therein. Here we summarize its characteristics
especially regarding the radiation hardness.

MND02 is equipped with four identical chains that process the signals from CCDs simultaneously.
Each chain consists of a preamplifier, a 5-bit digital-to-analog
converter (DAC), and two $\Delta\Sigma$ modulators (see Fig.~$\!$1 in \cite{Nakajima11}).
Analog signals from CCD are fed into MND02 through AC coupling capacitor.
Preamplifier differentially 10~times amplifies the signals and the 5-bit DAC
gives offset to the signal level. Two $\Delta\Sigma$ modulators work by turns
converting the amplified analog signals into two 155-bit serial streams \cite{Doty06}.
Then the bit stream is decimation-filtered in subsequent circuits implemented
in another ICs such as field-programmable gate arrays (FPGAs) and finally we
obtain the 12-bit decimal value.
The weighting coefficients for each bit in the decimation filter has been determined
by simulations (Fi.g~$\!$6 in \citep{Matsuura07}) to enhance
the frequency response as a low-pass filter and improve signal-to-noise ratio.

\begin{figure}
 \begin{center}
  \begin{tabular}{c}
  \hspace{-5mm}
   \includegraphics[width=0.5\textwidth]{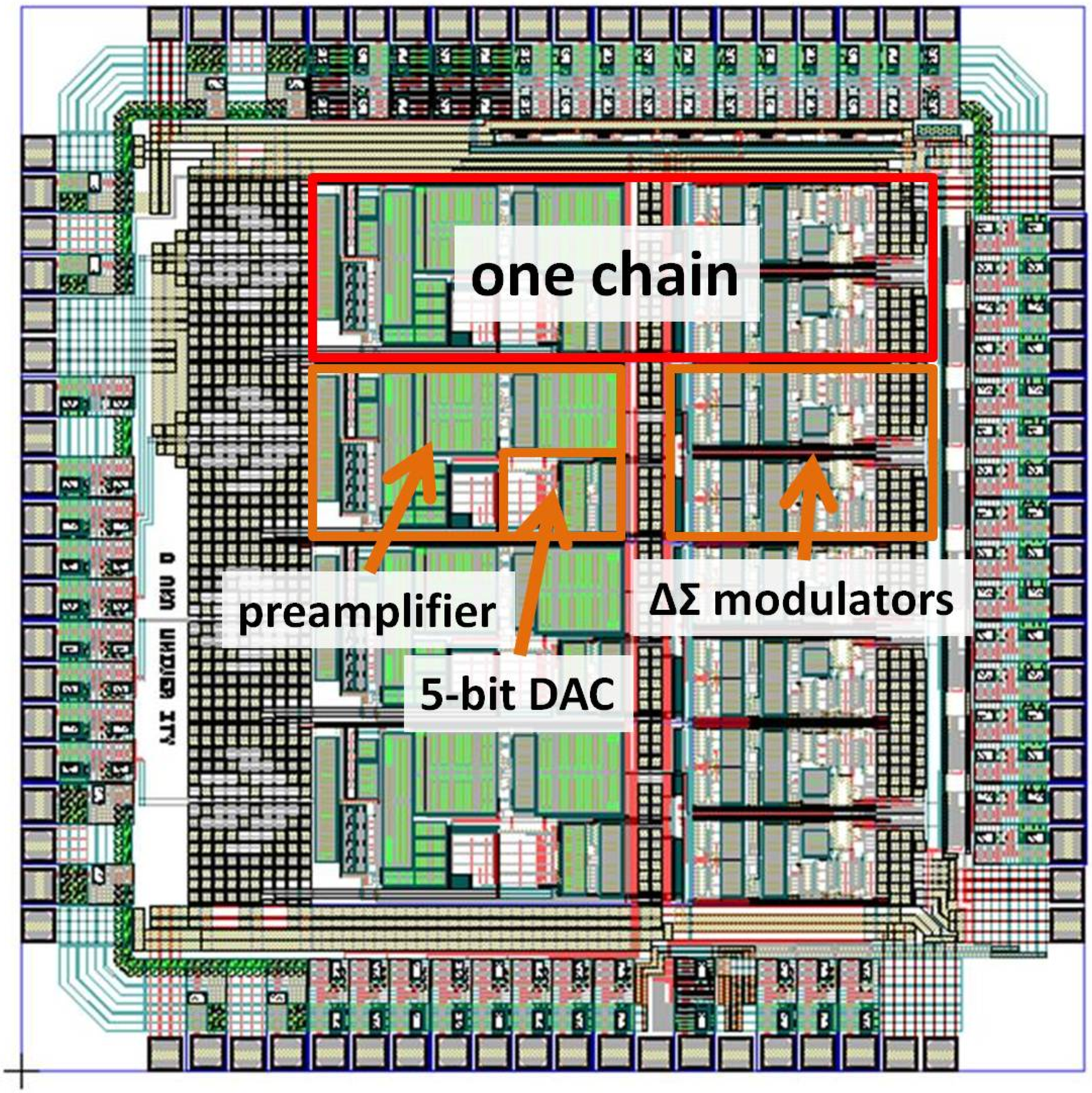}
  \end{tabular}
 \end{center}
 \caption[example] 
   { \label{fig:mnd02masklayout} 
Mask layout of MND02 illustrating the location of the circuit components.
The red box surrounds the region of single chain. Four identical chains are located lengthwise.
The orange boxes indicate the location of the circuit components in a chain.
The size of the chip is 3~$\!$mm square.
}
\end{figure}

It was fabricated by Taiwan Semiconductor Manufacturing Company (TSMC)
0.35~$\mu$m CMOS process through MOSIS service.
The chip wafer is made from a p-type epitaxial wafer to improve the SEL tolerance.
Although we do not employ the enclosed layout transistors, all the MOS toransistors
are surrounded by guard rings to strengthen the chip against TID effect.
3~mm square bare chip is packed into 15~mm square ceramic quad flat package (CQFP).
It functions with 3.3~$\!$V single power supply for analog and digital circuits.
Fig.~$\!$\ref{fig:mnd02masklayout} shows the mask layout of MND02 with the
labels of the circuit components. Most of the wafer area is devoted to capacitors.


\subsection{Specification of the heavy ion beam}
\label{ssec:beam}


We employed the charged-particle beams at a physics-general experiment 
line 1 (PH1) and the medium energy beam (MEXP) course at
Heavy Ion Medical Accelerator at Chiba (HIMAC) in National Institute for Radiological
Sciences.
Particles are extracted into the MEXP course immediately after
the linear accelerator. The particle kinetic energy per atomic mass unit is 6~MeV/u.
The beam duration is 350~$\!$$\mu$sec and can be shortened by a chopper
down to 5~$\!$$\mu$sec. The pulse interval is 1.65~sec and hence the time profile
of the beam is quite intermittent. The beam intensity is measured by the Faraday
cup (FC) located at the upstream of our chamber. The device under test (DUT)
is put in a vacuum state when we use MEXP course to avoid the energy loss
and the scattering of the incident particles.

In the case we use PH1 course, the particles are accelerated in the
synchrotron ring up to 400~MeV/u before they are extracted.
Since the particles are fed to two rings by turns, the beam interval is 3.3~$\!$sec,
while the duration of the beam on phase is  1~$\!$sec at the beam exit. 
DUTs are in an atmospheric pressure in PH1 course and are put 30~$\!$cm downstream
from the beam exit that consists of Aluminum film with the thickness of 100~$\!$$\mu$m.
The energy loss in the flim and air is well below 0.1\% of the incident energy.
Beam fluxes were measured using a scintillation counter prior to each test.

Table~\ref{tab:beam_spec} lists the specification of the beams adopted in the HIMAC.
LET ranges from 5.89$\times$10$^{-3}$ to 57.9~$\!$MeV$\cdot$cm$^2$/mg.
The beam size was measured by FC located near beam exit and the incident fluxes was
derived by fitting the beam profile with a two-dimensional Gaussian model.
There is no resin cover above the bare chip and the lid of the CQFP was removed throughout
the test to ensure the incident energy. Since the size of the socket on which DUTs are mounted
is 50~$\!$mm by 60~$\!$mm, the other ICs on the printed circuit board (PCB)
were not exposed to the radiation.


We supply 3.3~$\!$V directly from transistor DC power supply unit (PSU) through
no serial resistors. Hence the SEL causes immediate current increase up to the limit value.
The current in the PCB during test and that of the limit are 0.13 and 0.2~$\!$A, respectively.
The PSU was programed to shut down immediately when it noticed the current
limit and supply voltage again after 5 seconds. We provided pseudo CCD signals into the DUT
with a pixel rate of 78~$\!$kHz during the test.

\begin{table*}[htb]
\caption{Specification of the beams at HIMAC.} 
\label{tab:beam_spec}
\begin{center}       
\begin{tabular}{lccr@{.}lcc}
\hline
\rule[-1ex]{0pt}{3.5ex}  Species &  Beam & Energy        & \multicolumn{2}{c}{Linear energy transfer} & Beam Width & Maximum intensity \\
\rule[-1ex]{0pt}{3.5ex}              &  course & (MeV/u)  & \multicolumn{2}{c}{(MeV$\cdot$cm$^2$/mg)}  & (mm in FWHM)    & (Ion/sec/cm$^2$) \\
\hline
\rule[-1ex]{0pt}{3.5ex}  Proton  & PH1 & 100 & \hspace{1.2cm} 5&89$\times$10$^{-3}$   & 3.8 $\times$ 1.1 & 1.7$\times$10$^9$ \\
\rule[-1ex]{0pt}{3.5ex}  Silicon  & PH1 & 400 & 0&49                             &  2.9 $\times$ 2.6 &  3.3$\times$10$^4$ \\
\rule[-1ex]{0pt}{3.5ex}  Krypton & PH1 & 200 & 4&72                             &  4.4 $\times$ 3.8 & 3.3$\times$10$^5$ \\
\rule[-1ex]{0pt}{3.5ex}  Iron           & PH1 & 400 & 1&68                             &  7.1 $\times$ 5.7 & 3.3$\times$10$^4$ \\
\rule[-1ex]{0pt}{3.5ex}  Xenon & MEXP & 6      & 57&9                         &  2.4 $\times$ 1.8 & 1.0$\times$10$^8$ \\
\rule[-1ex]{0pt}{3.5ex}            & PH1 & 200 & 10&6                              &  diameter of 10 & 3.3$\times$10$^6$ \\
\rule[-1ex]{0pt}{3.5ex}            & PH1 & 400 & 7&2                               & diameter of 10 &  3.3$\times$10$^4$ \\
\hline
\end{tabular}
\end{center}
\end{table*} 

\section{Results of SEE Tests}
\label{sec:results}

\subsection{Results of SEL Tests}
\label{ssec:selresults}

\begin{figure}
 \begin{center}
  \begin{tabular}{c}
  \hspace{-5mm}
   \includegraphics[width=0.5\textwidth]{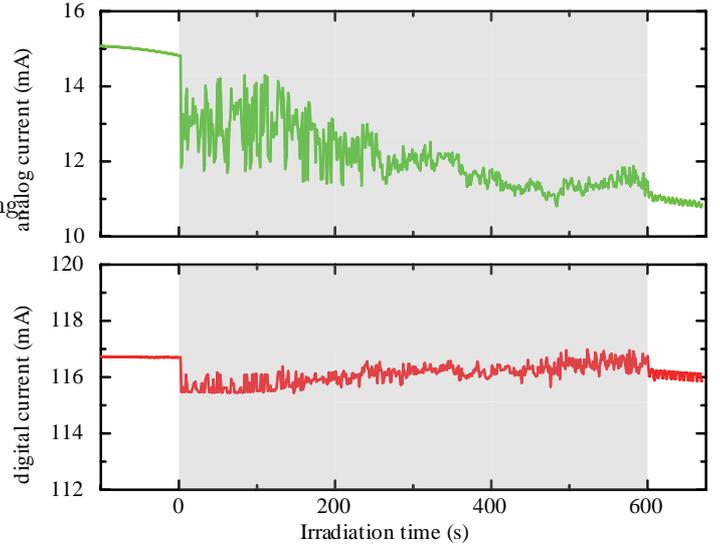}
  \end{tabular}
 \end{center}
 \caption[example] 
   { \label{fig:pcbcurrentsel} 
Time profile of the current in the PCB for analog circuit (top panel) and digital circuit
(bottom panel) during the beam irradiation
of Xenon 6~$\!$MeV/u (hatched duration). The current limit of the PSU
is set to be 0.2~$\!$A for the total of analog and digital circuits.
}
\end{figure}

Fig.~$\!$\ref{fig:pcbcurrentsel} shows the time profile
of the current in the PCB for analog and digital circuit in the case of Xenon beam.
The signal processing was performed throughout the duration shown in the figure.
The current level during the beam on does not increase compared
with that during beam off, which means there is no SEL during
the irradiation. Total fluence of the Xenon ions was 7.2$\times$10$^{10}$~$\!$/cm$^2$.

The cross-section of the SEL ($\sigma_{\rm SEL}$) is calculated dividing the number
of the SEL events by the fluence of the particles. Since there was no SEL event in this experiment,
we estimate the upper limit of $\sigma_{\rm SEL}$ using Poisson statistics. Assuming three SEL
events for the above fluence, the derived $\sigma_{\rm SEL}$ is below
$4.2\times10^{-11}$~$\!$cm$^2$/(Ion$\times$ASIC) at 95~$\!$\% confidence level.
\citet{Nakajima11} also performed the SEL test for MND02
but only for Fe ions with the LET of 1.68~MeV$\cdot$cm$^2$/mg. In this report we
put the upper limit of $\sigma_{\rm SEL}$ at
an LET by 34~times higher than that of the previous experiment. This LET is sufficiently high
enough that SEL event should not be a cause of instrument downtime.

Although there is no significant increase of the current due to the beam, some features
are seen in Fig.~$\!$\ref{fig:pcbcurrentsel}. The fluctuations of the current seen only when
the beam was on can be understood that it is due to the ion-shunt \citep{Hauser85,Knudson86,Brown93}.
It often follows from funneling \citep{Hsieh81,Campbell85} in the drain depletion layer.
When the carriers created by the charged particle concentrate throughout a volume linking two junctions,
the coupling of the junctions occurs and then the carriers redistribute. This results in the temporal increase
of the current in the chip.
On the other hand, the current jump down when the beam sets in and the small fluctuations after
the beam off cannot be explained by the ion shunt. Some charge up in the chip may explain the
former phenomenon.
Aside from the fluctuations, there is a slope in the current versus time. This comes from the other ICs
than MND02 on the PCB since we see this slope not only in the radiation experiments but also in other
experiments.

\subsection{Results of SEU Tests}
\label{ssec:seuresults}

\begin{figure}
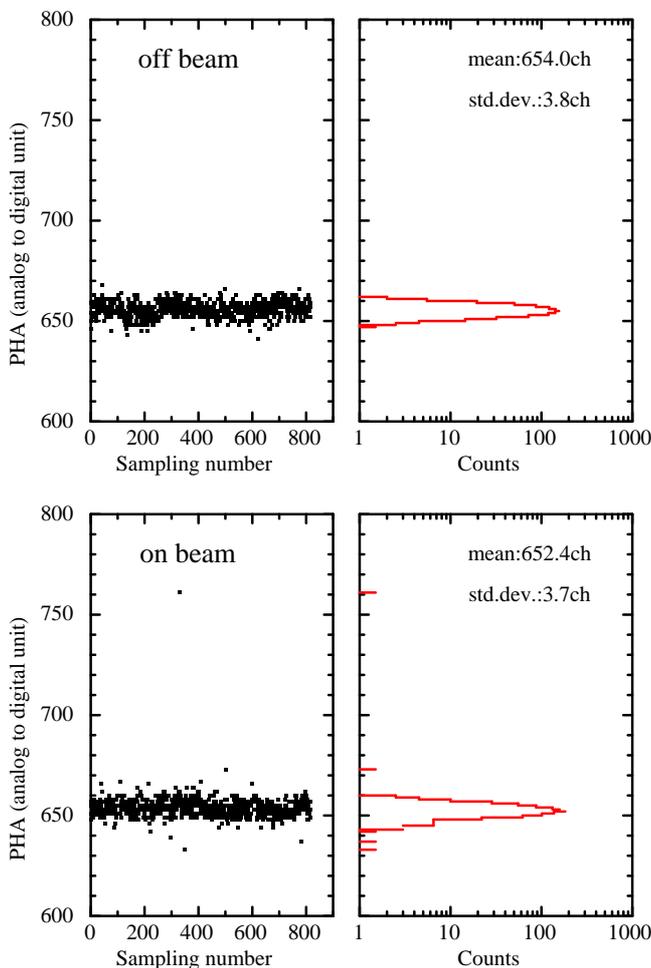

 \begin{center}
  \begin{tabular}{c}
  \hspace{-5mm}
   \includegraphics[height=0.5\textwidth, angle=-90]{SEU_type1_OFF_rev.ps} \\
  \hspace{-5mm}
   \includegraphics[height=0.5\textwidth, angle=-90]{SEU_type1_ON_rev.ps}
  \end{tabular}
 \end{center}
 \caption[example] 
   { \label{fig:scatterplot} 
Top panels: PHA distribution of the decimated pseudo 820 signals obtained during beam off
and projected histogram of their PHA.
Bottom panels: The same as top panels but for the data obtained during the beam irradiation.
Several data points show deviated PHA. 
}
\end{figure} 

Fig.~$\!$\ref{fig:scatterplot} shows the pulse height amplitude (PHA) distributions obtained
when we process the
pseudo CCD signals of 820~pixels with constant input voltage. While the distribution with
beam off can be reproduced with Gaussian, there are some anomalous pixels whose PHA
show significantly deviated values from the Gaussian when the Xenon 200~MeV/u beam
is on. We regarded the pixels that deviates more than 4~$\sigma$ from the distribution
center as SEU events.

The influence of the SEU events on the PHA distribution varies considerably
among the circuit components in MND02. If the bit inversion occurs at the resistor in the 5-bit DAC,
all the pixels will be affected after the impingement until we set the DAC value again.
Assuming that the digital circuit in the final stage of $\Delta\Sigma$ modulator is hit,
one of the 155 output bit stream is inverted. Then the expected deviation from the
normal PHA depends on the weighting coefficient of the inverted bit and it is no more
than 8~$\!$\% of the normal value \citep{Matsuura07}, which corresponds to 50~$\!$units in the case of
Fig.~\ref{fig:scatterplot}. In cases where big amounts of electron-hole pairs are
created in the capacitors of the preamplifier or the integrator in the
$\Delta\Sigma$ modulators, the amount of the PHA deviation from the normal value
depends on the energy deposited in the chip. Then only the single pixel is affected
since we assert a reset signal to both of the preamplifier and the $\Delta\Sigma$ modulator.
Considering the fact that the amount of the PHA deviation from the distribution ranges from
20 to 120~$\!$units and the fact that the anomalies do not last over pixels, we guess that the
probable components affected are the capacitors lying scattered over the chip.

\section{Estimation of the SEE Tolerance}
\label{sec:discussion}

\subsection{LET spectrum in the LEO}
\label{ssec:let_spectrum}

\begin{figure}
 \begin{center}
  \begin{tabular}{c}
  \hspace{-5mm}
   \includegraphics[height=0.5\textwidth, angle=-90]{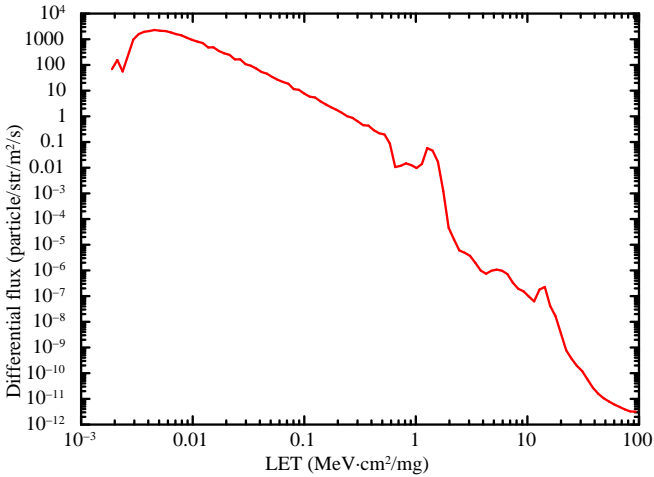}
  \end{tabular}
 \end{center}
 \caption[example] 
   { \label{fig:letspectrum} 
LET spectrum expected in the environment MND02 is located. Assumed orbital
parameters are the height of 550~$\!$km, the inclination angle of 30~$\!^{\circ}$.
Aluminum cover with the sickness of 20~$\!$mm surrounds MND02 for all direction.
The calculated duration is one year from January 2014 and an influence of
magnatic storm is taking into account.
}
\end{figure}

The SEE event rate of an electronic devices in the specific orbit and duration is estimated
by integrating the product of the LET flux and SEE cross-section.
The energies, densities and types of particles in the space environment depend significantly
on the orbital parameters such as the altitude, the inclination angle, the recent solar activity, and
the amount of spacecraft shielding.
Then we simulated the LET flux distribution due to the galactic cosmic-rays in the LEO
of our primary target mission ASTRO-H using the tools provided by JAXA Space Environment
\& Effects System (SEES) group \footnote[1]{http://seesproxy.tksc.jaxa.jp}, which is shown
in Fig.~$\!$\ref{fig:letspectrum}. In this simulation
Cosmic Ray on Micro-Electronics (CREME) code \citep{Adam81} is used.
In this mission MND02 chips are mounted on a PCB in the camera body that is
made of Aluminum with the thickness of about 20~$\!$mm.

\subsection{Estimation of SEE rate}
\label{ssec:see_rate_estimation}

The probability of SEE when a charged particle impact the LSI is expressed as a function
of LET ($L$) as following equation called Weibull curve.

\begin{equation}
P_{\rm SEE}(L) = P_0 \times (1 - {\rm exp}(-\frac{L-L_{\rm th}}{W}))
\label{eq:weibull}
\end{equation}

where $P_0$ is the saturated probability, $L_{\rm th}$ is the threshold LET,
and $W$ is a curve width. Then we calculate the SEE rate integrating the product
of $P_{\rm SEE}(L)$ and particle flux throughout the available LET range in Fig.~$\!$\ref{fig:letspectrum}
of 2.0$\times10^{-3} \le L \le$ 100~$\!$MeV$\cdot$cm$^2$/mg.

Although series resistors are implemented in power lines for the flight model electronics
to avoid the thermal destruction in the case of the SEL, we require the probability for the latch-up
to be below once per 30~$\!$yrs, which is 10 times the required mission lifetime of ASTRO-H.
Since we put the upper limit of the SEL probability
using Xenon beam of 6~$\!$MeV/u, the most pessimistic estimation of SEL
rate is $P_{\rm SEL} = 4.2\times10^{-11}$~$\!$cm$^2$/(Ion$\times$ASIC) for
2.0$\times10^{-3} \le L \le$ 57.9~$\!$MeV$\cdot$cm$^2$/mg and  $P_{\rm SEL} = 1$ for
57.9$\le L \le$ 100~$\!$MeV$\cdot$cm$^2$/mg. Then the SEL rate is calculated to
be once per 1.5$\times10^{9}$~$\!$sec or 49~$\!$yrs, which satisfies the requirement.

The SEU rate can be estimated in the same manner as that for SEL. However,
we need an additional care about the proton data because of the nuclear reactions
between high-energy protons and silicon nuclei in the device.
Protons indirectly induce SEUs since their elastic and inelastic interactions create
secondary projectiles. They consist of neutrons, alpha particles and heavy recoiling
ions \citep{Akkerman96}. The resultant LET distribution in the silicon wafer is not a $\delta$-function
but wide-spread function.

Hence we desire to calculate the number of SEU events due to the secondary
heavy ions and subtract it from the total number of SEUs in the proton test.
Since we measured the SEU probability
at the LET of Si 400~$\!$MeV/u ($L_{\rm Si400}$), it is assumed that
the secondary particles with the LET larger than $L_{\rm Si400}$ induce SEUs with the same
probability as $P_{\rm SEU}(L_{\rm Si400})$. On the other hand, when the LET of the secondary particle is
smaller than $L_{\rm Si400}$, we regard the SEU as that due to the proton.
We adopted the LET distribution calculated by \citet{Barak01} in which 
4$\times$10$^{-5}$ of the total number of protons create the secondary heavy ions whose
LETs are larger than $L_{\rm Si400}$.

\begin{figure}
 \begin{center}
  \begin{tabular}{c}
  \hspace{-5mm}
   \includegraphics[height=0.5\textwidth,angle=-90]{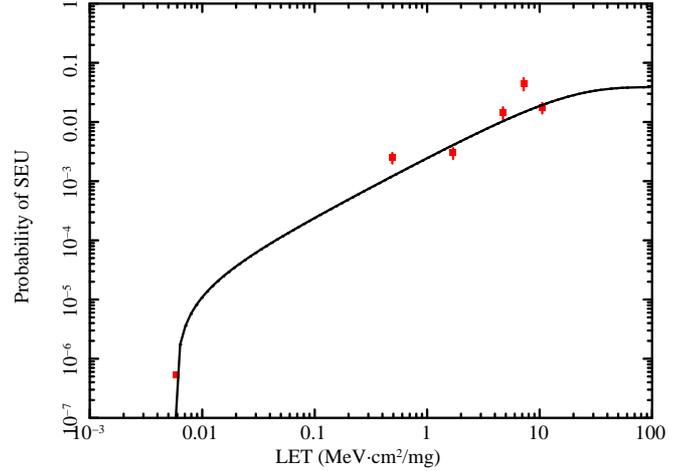}
  \end{tabular}
 \end{center}
 \caption[example] 
   { \label{fig:seudataweibull} 
Probability for SEU to occur as a function of LET. The results of 3 DUTs are averaged for each LET,
except that we tested 4 chips for proton beam.
The recoil component is considered for protons (5.89$\times$10$^{-3}$~$\!$MeV$\cdot$cm$^2$/mg).
Solid line is the Weibull curve fitted with the data.
}
\end{figure} 

Taking into account the above consideration, we fitted the SEU probability plot with the Weibull function
(Fig.~$\!$\ref{fig:seudataweibull}).
The best fit values are 5.68$^{+0.05}_{-0.02}\times$10$^{-3}$~$\!$MeV$\cdot$cm$^2$/mg for $L_{\rm th}$,
3.886$\pm0.001\times$10$^{-2}$ for $P_0$, and 15.3$^{+166.6}_{-9.7}$~$\!$MeV$\cdot$cm$^2$/mg for $W$.
We then estimated the SEU event rate using fitted parameter to be 8.0$^{+13.9}_{-7.3}\times$10$^{-6}$ ~$\!$SEU/sec.

Nevertheless the sharpness of the rising edge in Weibull function strongly depends on
this estimation, the number of the data below 1~$\!$MeV$\cdot$cm$^2$/mg is limited
in our experiment. Therefore we estimate the systematic error as follows considering that
the Weibull function is monotonically increasing. In the worst case estimation, the probability of SEU
in an LET range higher than a data point and lower than the next one is assumed
to be that of the latter data point. In this way we estimated
most pessimistic rate to be 1.3$\times10^{-3}$~$\!$SEU/sec.

Most of the SEU events such as that in Fig.~\ref{fig:scatterplot} cannot be distinguished
from the signals due to X-ray photons from astronomical objects. This means we need to
consider the SEU event as a part of the non X-ray background (NXB). The NXB intensity
of the X-ray CCD Camera onboard ASTRO-H is expected to be similar to that of the
X-ray Imaging Spectrometer onboard Suzaku satellite, which is 0.1~$\!$counts/sec \citep{Koyama07}.
Since the upper limit of the SEU rate is below 1.3~$\!$\% of known NXB rate,
it should not be a major component of the instrumental background.

\section{Conclusion}
\label{sec:summary}
We have developed an mixed-signal ASIC for the readout circuit of the onboard CCD camera.
The results of the radiation tolerance test for SEE using protons and heavy ion beams
can be summarized as follows

\begin{enumerate}

\item We detected no SEL events even with a high LET beam of Xenon 6~MeV/u.
The upper limit of the cross-section is $\sigma_{\rm SEL}$ $< 4.2\times10^{-11}$~$\!$cm$^2$/(Ion$\times$ASIC).
The flapping of the current seen only when the beam was on can be
understood such that it is due to the ion-shunt.

\item Some anomalous pixels that have peculiar PHA are seen only when the
beam is on, which we regard as SEU events. The most probable location of the impingement
is capacitors in the preamplifier and the integrator of the  $\Delta\Sigma$ modulators.

\item The SEE event rate in an LEO is estimated by integrating the product of
the simulated LET flux spectrum and SEE probability. The upper limit of the SEL rate
is once per 49 years, which satisfies our requirement.
The upper limit of the SEU rate is
derived to be 1.3$\times10^{-3}$~$\!$events/sec, which is 1.3~$\!$\% of known NXB rate.

\end{enumerate}


\section*{Acknowledgment}

We acknowledge people of Accelerator Engineering Corporation, especially Satoshi Kai who
offered us much supports in the radiation damage experiment. We used
the simulation tools provided by SEES group in JAXA.
This work is supported by the Nano-Satellite Research and Development Project in Japan
, the Research Project with Heavy Ions at NIRS-HIMAC, and the JSPS KAKENHI Grant Number
22740122, 23000004, and 24684010.

\end{document}